# MediaSpeech: Multilanguage ASR Benchmark and Dataset

*Rostislav Kolobov[1, 2], Olga Okhapkina[3], Olga Omelchishina[a4], Andrey Platunov[2], Roman Bedyakin[3], Vyacheslav Moshkin[3], Dmitry Menshikov[2], Nikolay Mikhaylovskiy[2, 5]*

[1]Tomsk Polytechnic University, Tomsk, Russia
[2]NTR Labs, Moscow, Russia
[3]AO HTSTS, Moscow, Russia
[4]Higher School of Economics – State University, Moscow, Russia
[5]Tomsk State University, Tomsk, Russia

nickm@ntr.ai

## Abstract

The performance of automated speech recognition (ASR) systems is well known to differ for varied application domains. At the same time, vendors and research groups typically report ASR quality results either for limited use simplistic domains (audiobooks, TED talks), or proprietary datasets. To fill this gap, we provide an open-source 10-hour ASR system evaluation dataset NTR MediaSpeech for 4 languages: Spanish, French, Turkish and Arabic. The dataset was collected from the official youtube channels of media in the respective languages, and manually transcribed. We estimate that the WER of the dataset is under 5%.

We have benchmarked many ASR systems available both commercially and freely, and provide the benchmark results.

We also open-source baseline QuartzNet models for each language.

## 1. Introduction

The availability of ImageNet [1] as an open and large image dataset have played a critical role for the development of deep learning paradigm. In the area of speech recognition, LibriSpeech [2] and AISHELL-1 [3] have probably played a similar role, providing ASR corpora widely used by both research and industry communities.

The performance of ASR systems is well known to differ for different application domains. Some researchers and practitioners call for further research democratization by making public more datasets from different application domains [4]. With the exploding popularity of semi-supervised learning [5][6][7][8][9] and availability of open-source tools [10] to collect audio datasets from YouTube providing general purpose domain ASR training datasets may be an overkill in 2021. Still, new datasets keep appearing, some of them with a very significant positive impact on the research community [11][12][13].

At the same time, vendors and research groups typically report ASR quality results either for limited use simplistic domains (audiobooks), or proprietary datasets, because the datasets collected in an automated fashion typically have low transcription accuracy. The problem is especially acute for non- {English, Mandarin} languages. To fill this gap, we provide an open-source 10-hour ASR system evaluation dataset for 4 languages: Spanish, French, Turkish, and Arabic, and benchmark various ASR systems against it.

We also open-source baseline QuartzNet [14] models for each language.

The dataset and models are available at https://github.com/NTRLab/MediaSpeech.

## 2. Data processing pipeline

### 2.1. Candidate audio selection

For the MediaSpeech dataset construction, we have selected a list of media with sufficient YouTube presence in each of the languages (see Table 1).

Table 1: *The list of media used*

| Language | Channel name | Language | Channel name |
|---|---|---|---|
| AR | Al Arabiya | FR | RT France |
| | FRANCE 24 Arabic | | France 24 |
| | BBC News عربي | | Russia Today |
| ES | Euronews | TR | Euronews |
| | BBC World | | FOX Haber |
| | CNN International | | Show Ana Haber |
| | Russia Today | | |

### 2.2. Downloading

For each of the channels listed in Table 1, video recordings containing speech were selected for a total of 20 hours per language. Audio tracks were loaded from the selected videos using the youtube_dl tool [15] library. Each audio track was converted to single-channel 16 kHz 16-bit PCM encoded WAV files using the FFmpeg library [16]

---
[a] Work performed during internship at NTR Labs

### 2.3. Audio segmentation

Segmentation of audio data was performed using the Vosk speech recognition toolkit [17]. Namely, Vosk was used to extract timestamps for each recognized word. Based on these timestamps, we sliced the audio track into segments less than 15 seconds long. Utterances that contain segments with no speech longer than 4 seconds were removed.

### 2.4. Transcription

A system for manual transcription has been built specifically for the project.

Each utterance has been first transcribed by an open-source ASR [17]. The transcription was used as a prompt for human transcribers to speed up transcription. Later analysis revealed that some transcribers just copied the automated transcription for more complex audio fragments. These utterances have been removed on the post-processing stage.

For each human transcriber, a transcription pipeline is built by the transcription system. For the quality control purposes, 5% of the utterances were taken from an existing spoken corpus (Mozilla Common Voice [18]). While the error rates of transcribers do not generalize to the new speech domain, control over these values allowed us to manage transcribers and make sure that the quality of transcription is consistent over time.

The initial attempts at crowdsourcing have shown that the transcription quality was inappropriate. Thus, we have hired full time at least three native speakers for each language. Table 2 lists WER for some transcribers (due to the bugs in the early versions of the transcription management system some data on the transcriber WER was lost).

Table 2: *Transcriber WER on CommonVoice corpus excerpts*

| Transcriber | Language | August | September | October | November |
|---|---|---|---|---|---|
| 1 | AR | 12.5 | 12 | 13 | 12.5 |
| 2 | AR | 10.5 | 11 | 9.5 | 10 |
| 3 | AR | 11 | 11.5 | 12.5 | 27 |
| 4 | TR | 10 | 11 | 10.5 | 9.5 |
| 5 | TR | 7 | 7 | 7 | 6 |
| 6 | TR | 11.5 | 7 | 7 | 7 |
| 7 | ES | 15 | 17 | 15.5 | 14.5 |
| 8 | ES | 11.5 | 11 | 10 | 10.5 |
| 9 | ES | 12 | 11 | 11.5 | 12 |
| 10 | FR | 13 | 14 | 15 | 14.5 |
| 11 | FR | 9.5 | 6 | 7 | - |

Each utterance has been transcribed by two human transcribers. In the case where the relative WER of transcriptions was over 5%, the third transcriber resolved the conflict.

All text files are encoded in UTF8.

### 2.5. Post-processing

Symbols such as <, >, [, ], ~, /, \, =, etc., are removed. Text normalization is applied towards numbers. Utterances containing URLs are removed. Abbreviations such as CEO are presented in lowercase.

### 2.6. Alphabet normalization

Some target languages, such as French and Spanish have alphabets that contain non-latin symbols such as "æ", "ü", "ă", etc. We have brought the alphabets used in the dataset into compliance with the modern polygraphic standards. The final alphabets are presented in Table 3.

Table 3: *Normalized Alphabets*

| Language | Alphabet |
|---|---|
| French | azertyuiopqsdfghjklmùwxcvbné'èçàêôâûœ |
| Spanish | abcdefghijklmnñopqrstuvwxyzáéíóúüé' |
| Arabic | أنت سير إلمحةاقثعهذفبئضودجصكخشزطءغظآؤ |
| Turkish | abcçdefgğhıijklmnoöprsştuüvyz' |

## 3. Speech recognition baseline

In our experiments, we finetuned English Quartznet 15x5 model [14]. We use a batch size of 16 per GPU on a single server with 4 T4 GPUs. All models were trained for 50 epochs with learning rate=0.005, using NovoGrad optimizer and SpecAugment data augmentation on proprietary in-domain datasets.

## 4. ASR comparison

### 4.1. ASRs tested

ASRs with the following models and parameters were tested on the dataset:

1. Deep Speech FR trained on a set of common voice [20]
2. Deep Speech ES [21]
3. Wav2Vec2 FR [22]
4. Wav2Vec2 AR [23]
5. Wav2Vec2 TR [24]
6. Wav2Vec2 ES [25]
7. Silero [26]
8. Wit [27]
9. Google Speech To Text [28] using BCP-47: es-ES, tr-TR, fr-FR, ar-BH
10. Azure "Speech" API [29] using BCP-47: es-ES, tr-TR, fr-FR, ar-BH
11. VOSK ES [30]
12. VOSK FR [31]
13. VOSK AR [32]
14. VOSK TR [33]

### 4.2. Benchmark results

Table 4 below lists Word Error Rates for each language and ASR system benchmarked.

Table 4: *WER benchmark*

| ASR | AR | FR | TR | ES |
|---|---|---|---|---|
| Azure | 0.3016 | **0.1683** | 0.2296 | 0.1292 |
| Google | 0.4464 | 0.2385 | 0.2707 | 0.2176 |
| VOSK | 0.3085 | 0.2111 | 0.3050 | 0.1970 |
| Silero | - | - | - | 0.3070 |
| Wit | 0.2333 | 0.1759 | **0.0768** | **0.0879** |
| Deepspeech | - | 0.4741 | - | 0.4236 |
| Quartznet (ours) | **0.1300** | 0.1915 | 0.1422 | 0.1826 |
| wav2vec2 | 0.9596 | 0.3113 | 0.5812 | 0.2469 |

### 4.3. Discussion

Unsurprisingly, the research ASR systems trained on our-of-domain data, like Common Voice (DeepSpeech, wav2vec 2.0) perform worse on the media domain than the commercial quality systems trained on diverse data.

Wit has somewhat surprisingly shown fantastic performance on our dataset. We assume a large portion of media in the Wit training dataset.

The multitude of Arabic dialects most likely accounts for low results on standard media Arabic.

## 5. Acknowledgements

The authors are grateful to colleagues at NTR Labs Machine Learning Research group and AO HTSTS for the discussions and support.